# Nonlinear unitary circuits for photonic neural networks

Sunkyu Yu[1]*, Xianji Piao[2]*, and Namkyoo Park[3]*

[1]Intelligent Wave Systems Laboratory, Department of Electrical and Computer Engineering, Seoul National University, Seoul 08826, Korea

[2]Wave Engineering Laboratory, School of Electrical and Computer Engineering, University of Seoul, Seoul 08826, Korea

[3]Photonic Systems Laboratory, Department of Electrical and Computer Engineering, Seoul National University, Seoul 08826, Korea

*Corresponding author. Email: sunkyu.yu@snu.ac.kr; piao@uos.ac.kr; nkpark@snu.ac.kr

**Abstract:** Photonics has unlocked the potential for energy-efficient acceleration of deep learning. Most approaches toward photonic deep learning have diligently reproduced traditional deep learning architectures using photonic platforms, separately implementing linear-optical matrix calculations and nonlinear activations via electro-optical conversion, optical nonlinearities, and signal-encoded materials. Here we propose a concept of nonlinear unitary photonic circuits to achieve the integration of linear and nonlinear expressivity essential for deep neural networks. We devise a building block for two-dimensional nonlinear unitary operations—featuring norm-preserving mappings with nonconservative inner products—which enables the construction of high-dimensional nonlinear unitary circuits. Using deep nonlinear unitary circuits, we demonstrate exponential growth in trajectory length and near-complete coverage of the output space, both of which are essential for deep learning. Along with neuroevolutionary learning examples for the regression of a nonconvex function, our results pave the way to photonic neural networks with highly expressive inference and stable training.

In sustaining the groundbreaking progress of deep learning [1], one urgent challenge is to develop hardware optimized for training and inference. Photonic hardware is a promising candidate for these tasks due to energy-efficient and ultrafast matrix calculations achievable with light [2]. Various approaches have demonstrated the potential for accelerating deep learning with superior energy efficiency, by employing integrated circuits [3], diffractive multilayers [4], and scattering systems [5]. In these platforms, weight matrices can be implemented in the realm of linear optics in spatial [3-7], temporal [8], and frequency [9] domains.

In contrast to linear-optical matrix calculations, achieving nonlinear expressivity in photonics remains a substantial challenge [2,10], despite its importance in deep neural networks (DNNs) [1]. Because Maxwell's equations can yield nonlinear responses only through materials, various light-matter interactions have been examined to express nonlinear mappings. For example, beyond simple photodetection that provides square-law nonlinearity [4], modulating optical components via detection-based feedback [11-13] and phase-change materials [14] enables richer forms of nonlinearity, such as ReLU-like [11-13] and memristic [14] responses. Recently, the nonlinear relationships between light and scatterers have been utilized to extract nonlinear expressivity within linear optics [5,15,16]. However, these methods diminish the advantages of photonic deep learning due to the need for considerable electronic signal processing.

Accordingly, despite their weak magnitudes, continuous efforts have been made to leverage optical nonlinearities to achieve nonlinear expressivity. Analogous to software activation functions [1] and corresponding electronic hardware [17], optical nonlinearities based on saturable absorption [3,18], cross-gain modulation [19], and quantum interference [20] have been employed to shape light intensity. However, altering light intensity involves both software and hardware costs: vanishing gradients during training and the need for loss compensation. Drawing on insights from isolated system assumptions in classical and quantum physics [21-23], and the success of unitary deep learning [24], one could envisage leveraging norm-preserving optical nonlinearity, such as self-phase modulation (SPM) [25], to achieve DNN expressivity.

Here, we propose nonlinear unitary (NU) circuits for norm-preserving photonic deep learning with enhanced expressivity, meeting the benchmark of software-implemented DNNs. By utilizing SPM in integrated photonic platforms, we implement two-dimensional (2D) NU gates to construct high-dimensional unitary circuits with programmable nonlinearity. Using deep NU circuits, we achieve norm-preserving state mappings with nonconservative inner products, yielding exponentially growing expressivity and near-complete state-space coverage as circuit depth increases—both essential for deep learning of advanced problems. The proposed circuit is validated through a neuroevolutionary regression example involving a nonconvex function. Our results highlight the norm-preserving integration of linear and nonlinear expressivity within a compact architecture, which is a desired configuration for both classical and quantum wave neural networks, inheriting the advantages of unitary neural networks [24].

**Nonlinear U(1) unit**

NU operations [26], which lead to norm-preserving mappings with non-conservative inner products due to nonlinearity, have been studied in nonlinear quantum mechanics for quantum walks [27], Bose-Einstein condensates [28], and squeezed states [29]. As a key component for applying high-dimensional and reconfigurable NU operations to photonic deep learning, we start from the design of an integrated photonic system that offers nonlinear U(1) group operations, referred to as NU(1). Figure 1a shows the NU(1) unit, which consists of a resonator with resonance frequency $\omega_0$, side-coupled to a waveguide with coupling lifetime $\tau$. This configuration enables the excitation of a counterclockwise travelling-wave resonance mode $\mu$ by the incident wave $\psi_+$. Assuming negligible intrinsic losses and backscattering, the unit achieves the outgoing wave $\psi_-$ with U(1) operations: phase-shifted perfect transmission [9,30] (Supplementary Note S1).

To introduce programmable nonlinearity to the unit, we leverage resonance perturbations, $\omega_0' = \omega_0 + \Delta\omega_L + \Delta\omega_{NL}$, where $\Delta\omega_L$ and $\Delta\omega_{NL}$ represent linear and nonlinear resonance shifts, respectively. The linear shift $\Delta\omega_L$ is tunable for programmability, achievable with electro-optic [31] or thermo-optic [32] phase shifters. When circuit pretraining is adopted [33], the required range of phase shifts can be reduced with adjusted resonator geometry. On the other hand, $\Delta\omega_{NL}$ is derived via SPM from the optical Kerr effect [25,34,35], which induces a red shift in the resonance, $\Delta\omega_{NL} = -\alpha|\mu|^2$ ($\alpha \geq 0$), where $\alpha$ is determined by the Kerr coefficient and the modal profile [25,35].

Using nonlinear temporal coupled mode theory [36], we establish the operation principles at steady state (Supplementary Note S2). First, when $|\Delta\omega_L| < \sqrt{3}/(2\tau)$ for the incidence frequency $\omega = \omega_0$, the nonlinear dynamics of the unit possesses a unique equilibrium. This condition leads to a deterministic phase shift without hysteresis, as $\psi_- = \exp(-i\xi)\psi_+$, where $\xi = \xi(|\psi_+|^2, \Delta\omega_L)$ represents the nonlinear function. Second, $\xi$ exhibits high nonlinear expressivity obtained through two mechanisms: the arctangent relationship between $\xi$ and the resonance perturbation, and the SPM-induced resonance perturbation determined by incident light. While the latter originates from nonlinear optics [25], the former arises from the linear-optical relationship between waves and matter [5,15,16].

Based on these principles, Figs. 1b and 1c illustrate the behavior of the unit at $\omega = \omega_0$. The results are verified by comparing steady-state analytical solutions and time-domain numerical solutions with the sixth-order Runge-Kutta (RK6) method [37]. When $\alpha = 0$ (Fig. 1b), the resulting phase shift $\xi$ is invariant against $|\psi_+|^2$, although the shift depends on $\Delta\omega_L$. In contrast, with nonzero $\alpha$ (Fig. 1c), the unit operates as a nonlinear phase shifter dependent on $|\psi_+|^2$, while the nonlinearity is programmable with $\Delta\omega_L$. Therefore, the proposed unit enables reconfigurable NU(1) operations (Supplementary Note S3 for temporal dynamics).

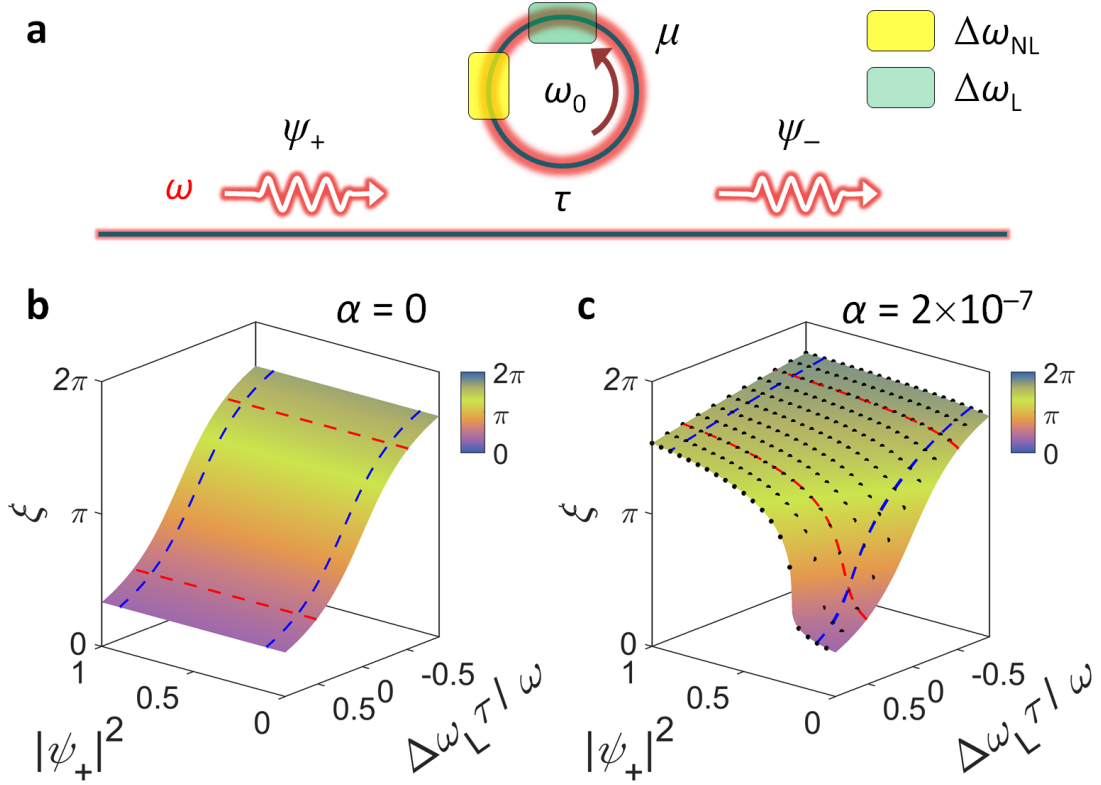

**Fig. 1. NU(1) photonic units**. **a,** Schematic of the unit. The boxes describe the setups for $\Delta\omega_L$ and $\Delta\omega_{NL}$. Although $\Delta\omega_{NL}$ is depicted as spatially confined for clarity, the effect occurs throughout the resonator when a homogeneous material is used. **b,c,** Phase shifts $\xi(|\psi_+|^2, \Delta\omega_L)$ in linear (**b**) and nonlinear (**c**) operations. The colored surfaces in (**b,c**) and black dots in (**c**) represent the analytical (Supplementary Note S2) and numerical (Supplementary Note S3) results, which are excellently matched. Blue dashed lines represent $|\psi_+|^2 = 0.1005$ and $0.8995$. Red dashed lines represent $\Delta\omega_L = \pm 0.5161\omega_0/\tau$. $\omega_0 = 1$ and $\tau = 2{,}000/\omega_0$.

## Nonlinear U(2) gate

Because the U(1) operation itself does not affect observables, we utilize NU(1) units to construct a nonlinear U(2) gate—NU(2) gate—which serves as a building block for NU circuits. Figure 2a illustrates the NU(2) gate, which consists of a Mach-Zehnder interferometer (MZI) followed by an NU(1) unit in each channel. The phase shifts $\xi^{(1,2)}$ in the upper (channel 1) and lower (channel 2) arms provide $2\pi$ coverage in their difference $\Delta\xi = \xi^{(1)} - \xi^{(2)}$.

The gate performs the norm-preserving operation $U = R_z(\Delta\xi) R_x(-\pi/2)$, where $R_x(\gamma)$ and $R_z(\gamma)$ denote $\gamma$-rotation operations about the $x$- and $z$-axes of the Bloch sphere, respectively. In $U(\Delta\xi)$, $\Delta\xi$ is the following nonlinear function:

$$\Delta\xi = \Delta\xi\left(\left[R_x\left(-\frac{\pi}{2}\right)\psi_+\right]^* \circ \left[R_x\left(-\frac{\pi}{2}\right)\psi_+\right], \begin{bmatrix}\Delta\omega_L^{(1)}\\\Delta\omega_L^{(2)}\end{bmatrix}\right), \quad (1)$$

where the symbols "∘" and "*" denote the element-wise Hadamard product and the complex conjugate, respectively, $\Delta\omega_L^{(1,2)}$ is the linear resonance shifts in channels 1 and 2, and $\psi_+$ is an input spinor state. Therefore, the gate executes an NU(2) operation dependent on the input, as $U$

= $U(\psi_+,\Delta\omega_L^{(1,2)})$, achieving norm preservation $\langle\psi_+|U(\psi_+)^\dagger U(\psi_+)|\psi_+\rangle = \langle\psi_+|\psi_+\rangle$ and nonconservative inner products $\langle\psi_+'|U(\psi_+')^\dagger U(\psi_+)|\psi_+\rangle \neq \langle\psi_+'|\psi_+\rangle$ when $\psi_+ \neq \psi_+'$. The variation in inner products is programmable through $\Delta\omega_L^{(1,2)}$ for $U(\psi_+,\Delta\omega_L^{(1,2)})$.

Figure 2b-d shows examples of the calculated gate operation. For the input states *A* and *B* (Fig. 2b), the MZI applies the ($-\pi/2$)-rotation about the *x*-axis, resulting in *A*' and *B*' (Fig. 2c). However, the magnitude of the subsequent *z*-axis rotation is greater in *B*' than *A*', because state *B*' supports higher optical intensity in channel 1. Consequently, the evolutions from *A* to *A*" and from *B* to *B*" differ, depending on the input (Fig. 2d).

Similar to linear circuits [6,7], more complex nonlinear circuits can be constructed using NU(2) gates. For example, higher-dimensional NU operations can be implemented by arranging the gates according to the Reck [38] or Clements [39] designs (Fig. 2e). However, we note that the NU operations of the circuits are not as straightforward as the analytically determined universal linear unitary operations [6-9,38-40].

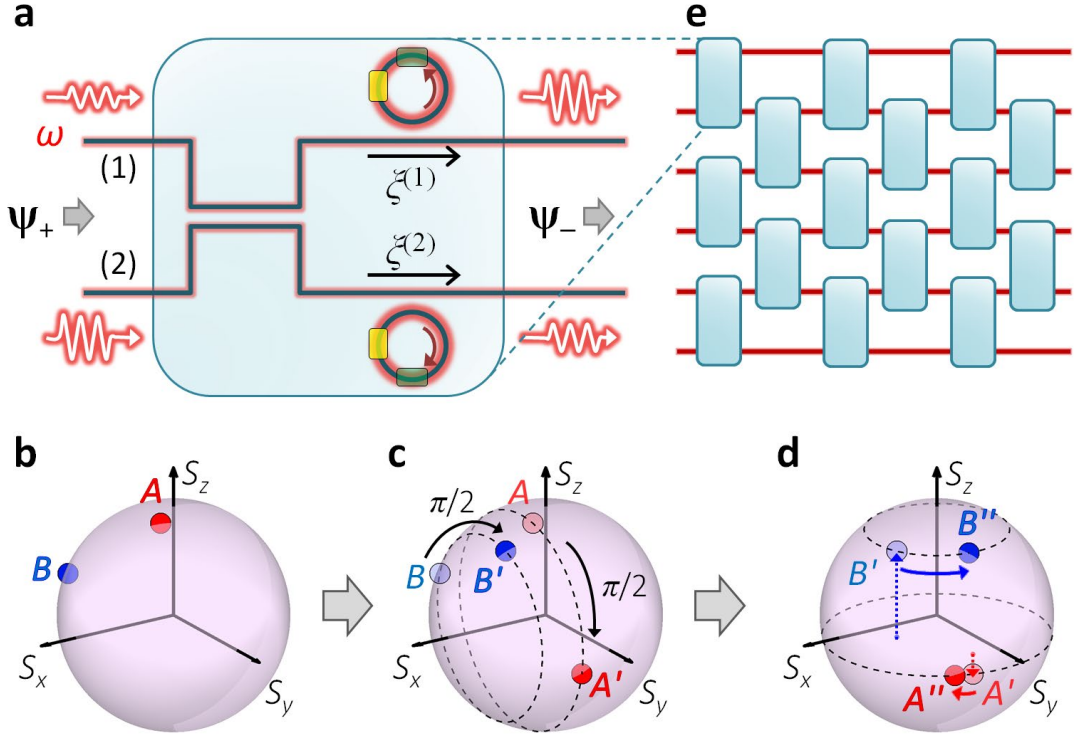

**Fig. 2. NU(2) photonic gates and circuits. a,** Schematic of the gate. The gate imposes an NU operation on $\psi_+$, resulting in $\psi_-$. **b-d,** Examples of the NU(2) operations: the input states *A* and *B* (**b**), the intermediate states *A*' and *B*' after the MZI (**c**), and the final states *A*" and *B*" (**d**). $S_{x,y,z}$ denote the spinor observables. The dashed arrows in **d** represent the $S_z$ component. All the parameters are the same as those in Fig. 1c. **e,** Schematic of an NU(6) circuit.

**Circuit expressivity**

To understand NU operations as a core process for photonic deep learning, we examine deep NU(2) circuits—the simplest form of deep NU DNNs. In particular, building on widely used benchmarks in conventional deep learning [41], we focus on the impact of varying circuit depth on circuit expressivity, which has mostly been addressed in a problem-specific and heuristic manner in prior studies on photonic DNNs [2-5,15,16].

Figure 3a illustrates an NU(2) circuit consisting of $M$ gates. The $m$-th gate performs a mapping between the spinor states $\psi_{m-1}$ and $\psi_m$ via the NU matrix $T_m$, which is determined by the input $\psi_{m-1}$ and the system parameters $\Delta\omega_{L,m}^{(1,2)}$ and $\alpha_m^{(1,2)}$. The NU mapping achieved with the circuit is given by $\psi_M = (\prod_{m=1}^{M} T_m)\psi_0$. The mapping corresponds to the geometrical transition from $(\theta_0,\varphi_0)$ to $(\theta_M,\varphi_M)$, where $(\theta_m,\varphi_m)$ denotes the spherical coordinates of the vectorial observables on the Bloch sphere, $\mathbf{S}_m = \langle\psi_m|\boldsymbol{\sigma}|\psi_m\rangle = [\sin\theta_m\cos\varphi_m, \sin\theta_m\sin\varphi_m, \cos\theta_m]^T$, for the Pauli vector $\boldsymbol{\sigma} = \mathbf{x}\sigma_x + \mathbf{y}\sigma_y + \mathbf{z}\sigma_z$ and the Pauli matrices $\sigma_{x,y,z}$.

Following the approach used in DNNs [41], we evaluate the expressivity of untrained circuits. Considering practical implementations with tunable phase shifters and a homogeneous nonlinear material, we define the system parameters using uniform and Bernoulli random distributions, as follows:

$$\Delta\omega_{L,m}^{(1,2)} \sim \text{Uniform}\left(\Delta\omega_0 - \frac{\sqrt{3}}{2\tau}, \Delta\omega_0 + \frac{\sqrt{3}}{2\tau}\right), \tag{2}$$

$$\alpha_m^{(1,2)} \sim \alpha_0 \times \text{Bernoulli}(0.5).$$

To investigate the impact of nonlinearity, we compare the expressivity of linear ($\alpha_0 = 0$ with $\psi_m = \psi_m^L$) and nonlinear ($\alpha_0 \neq 0$ with $\psi_m = \psi_m^{NL}$) unitary circuits.

We examine two key metrics—complexity and coverage—to quantify the expressivity of untrained circuits. First, compared to linear circuits [38,39], the primary advantage of the NU circuits is expected to lie in their enhanced nonlinear complexity, which is a crucial benchmark in deep learning **[1,41]**. This feature can be captured by trajectory length, which measures how the output changes as the input varies along a path in the input space [41]. Extending this quantity to the spinor representation, we define a metric for trajectory growth on the Bloch sphere, as follows:

$$\mathcal{E}(\{\psi_0(v)\}; v_i \leq v \leq v_f) = \frac{\int_{v_i}^{v_f} \left\|\frac{d\mathbf{S}_M}{dv}\right\| dv}{\int_{v_i}^{v_f} \left\|\frac{d\mathbf{S}_0}{dv}\right\| dv}, \tag{3}$$

where $\{\psi_0(v)\}$ represents the input state trajectory defined by the parameter $v$ ranging from $v_i$ to $v_f$. We note that $\mathcal{E}$ quantifies the variation of the trajectory length through the circuit for a randomly shaped input trajectory $\{\psi_0(v)\}$ (Supplementary Note S4).

Figure 3b shows the numerically calculated trajectory variations through the circuits of different depths, comparing linear and nonlinear circuits. While linear circuits preserve the original geometry of the input trajectory, nonlinear circuits introduce substantial complexity with increased trajectory lengths, which originates from varying inner products due to optical nonlinearity. Remarkably, we observe exponential growth in trajectory length with circuit depth (Fig. 3c), which is a unique characteristic of DNNs [41] that underlies their powerful expressivity. Therefore, NU circuits endow DNNs with high expressivity, while retaining a key advantage: conservation of the entire signal level $\langle\psi_m|\psi_m\rangle$ throughout the network.

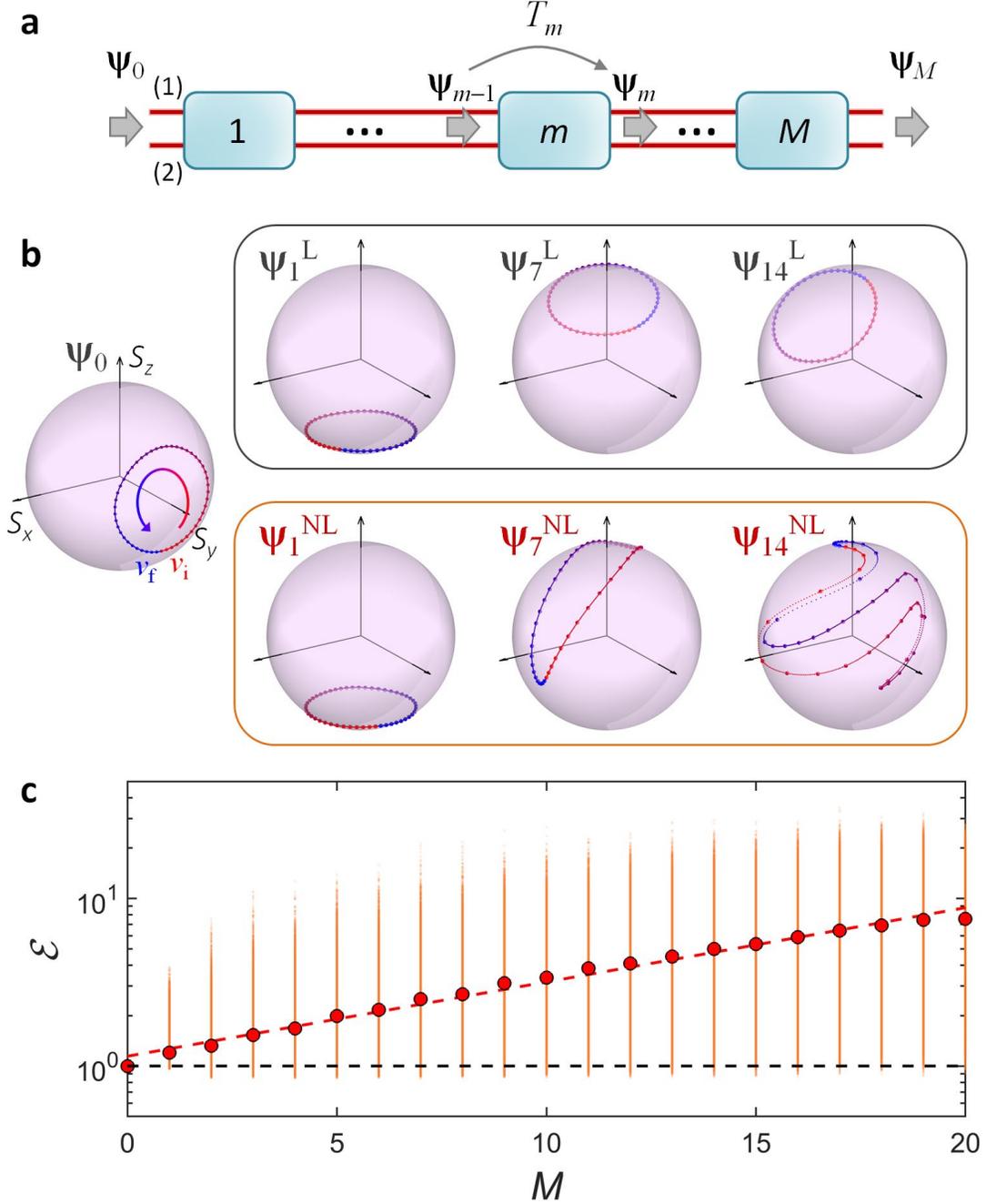

**Fig. 3. Mapping complexity**. **a,** Schematic of the NU(2) circuit used to examine expressivity. $T_m$ denotes the unitary operation achieved by the $m$-th gate. **b,** Examples of trajectory evolution through linear (top panel) and nonlinear (bottom panel) circuits for the same input trajectory (left panel) at $M = 1$, 7, and 14. The parameters $v_i$ and $v_f$ for Eq. (3) are indicated in red and blue, respectively. **c,** Trajectory growth as a function of $M$. Orange points represent $2\times10^4$ realizations: $10^3$ random system configurations according to Eq. (2), and 20 different input trajectories for each system. Red circles depict the ensemble average of $\mathcal{E}$. A black dashed line represents linear circuits: $\mathcal{E} = 1$. The steady-state solutions in Fig. 1 are used for **b** and **c**.

Although higher $\mathcal{E}$ provides greater complexity in mapping the input and output Hilbert spaces, ideal DNNs also require complete coverage of the output space for each input state during network training. Notably, traditional linear unitary circuits with universal SU(2) gates allow for complete coverage [38,39]. However, the introduction of nonlinearity, which causes trajectory deformation and state concentration in the output space due to varying inner products, no longer guarantees complete coverage.

To quantify coverage, we estimate the average range of the subspace of $\psi_M = (\prod_1^M T_m)\psi_0$ with respect to the output space—$(\theta,\varphi)$ on the Bloch sphere—for an arbitrary input $\psi_0$ and random untrained circuits. We define the coverage $\mathcal{C}$ for this estimation using normalized entropy, as follows:

$$\mathcal{C} = -\frac{1}{\log N_{\text{cell}}} \sum_{i,j} p_{ij} \log(p_{ij}), \tag{4}$$

where $N_{\text{cell}}$ is the number of the 2D cells discretizing the output $(\theta,\varphi)$, and $p_{ij}$ represents the probability that the output state is in the $(i,j)$-th cell (Supplementary Note S5). While $\mathcal{C}$ ranges from 0 to 1, $\mathcal{C} = 1$ represents complete coverage of the output space.

Figure 4a shows the numerically calculated output states for a specific input, illustrating the rapid increase in coverage with circuit depth. As demonstrated in Fig. 4b, the degradation of $\mathcal{C}$ due to nonlinearity becomes negligible for $M > 7$ with $|\langle\mathcal{C}_L\rangle_I - \langle\mathcal{C}_{NL}\rangle_I| < 10^{-3}$, where $\mathcal{C}_L$ and $\mathcal{C}_{NL}$ represent the coverages of the linear and nonlinear circuits, respectively, and $\langle\ldots\rangle_I$ denotes the ensemble average over input states. Therefore, deep NU circuits provide a highly expressive platform for DNNs, offering both complexity and coverage in the mapping of input and output spaces.

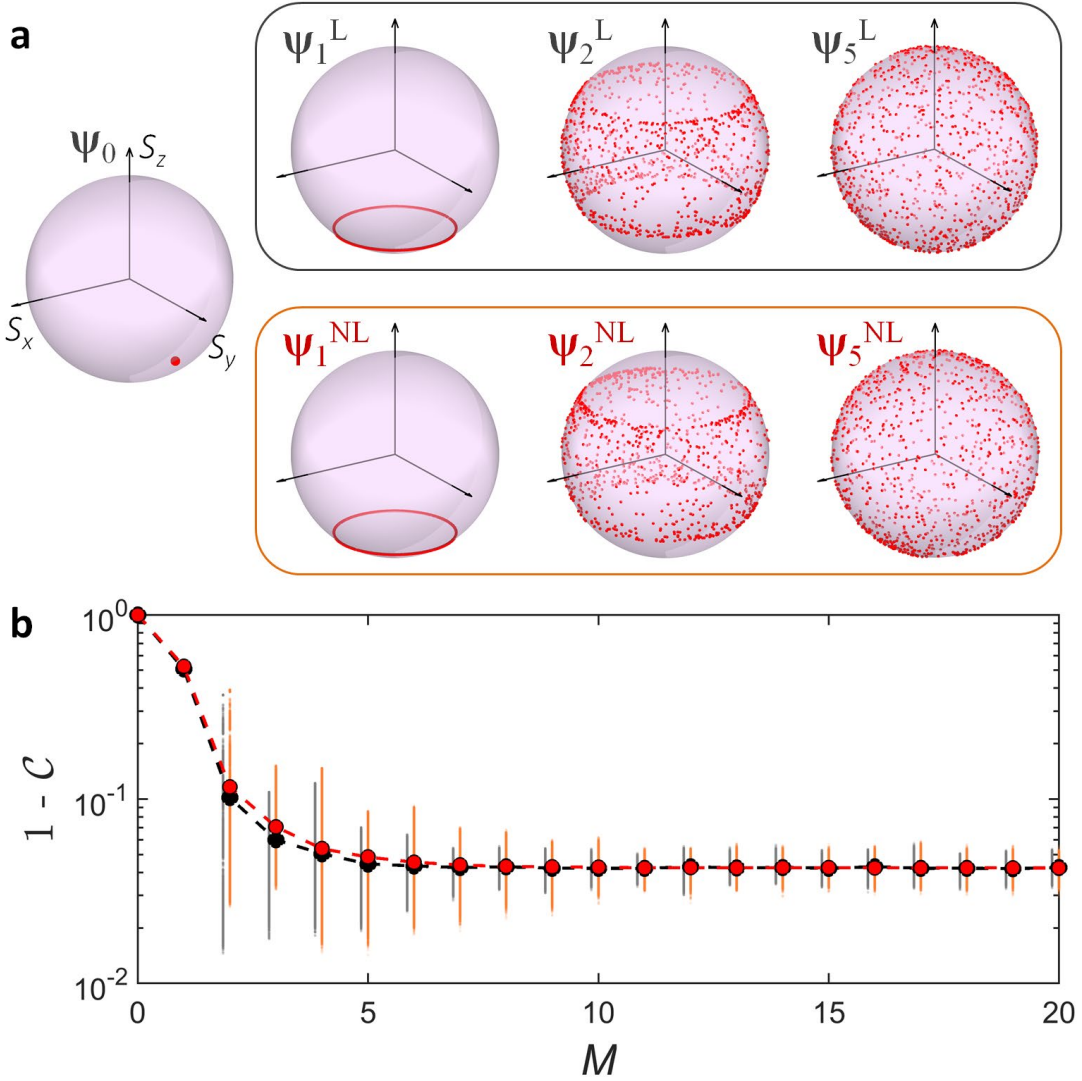

**Fig. 4. Mapping coverage**. **a,** Examples of resulting output states through linear (top panel) and nonlinear (bottom panel) circuits for an input state (left panel) at $M = 1, 2$, and 5. **b,** Coverage gap $1 - \mathcal{C}$ as a function of $M$ in linear (gray and black) and nonlinear (orange and red) circuits. Both gray and orange points represent $2 \times 10^4$ realizations from different inputs $\psi_0$: $100 \times 200$ discretization of the $\theta$–$\varphi$ input space. Both black and red circles depict the ensemble averages of $1 - \mathcal{C}$. The results of the linear circuits are slightly offset from integer $M$ values for clarity. The steady-state solutions in Fig. 1 are used.

**NU DNNs**

Based on the demonstrated expressivity of deep NU(2) circuits, we develop $M$-depth NU DNNs for a regression example. The target function is a 2D nonconvex Rastrigin function [42], with 2D inputs and a one-dimensional (1D) output encoded in the $(\theta_0,\varphi_0)$ coordinates of $\mathbf{S}_0$ and the $\theta_M$ coordinate of $\mathbf{S}_M$, respectively (Supplementary Note S6). After initializing the circuit with Eq. (2), we employ neuroevolutionary learning [43,44] combined with simulated annealing [45] to address the local minima issue in nonconvex regression (Supplementary Note S7). The total number of training parameters is $4M$: $2M$ continuous and $2M$ binary variables.

Figures 5a and 5b show the dependency of regression performance on circuit depth. The mean square error cost function $\rho$ is evaluated using 1,000 training data points in Fig. 5a and 200 test data points in Fig. 5b. As expected from Fig. 3c, the deeper NU circuit demonstrates superior performance. This observation is further confirmed with plots of 5,000 test data points, comparing the original function (Fig. 5c), the untrained network outputs (Fig. 5d), and the trained network outputs (Fig. 5e). The results illustrate that the increased complexity at higher $M$ values enables improved regression through training.

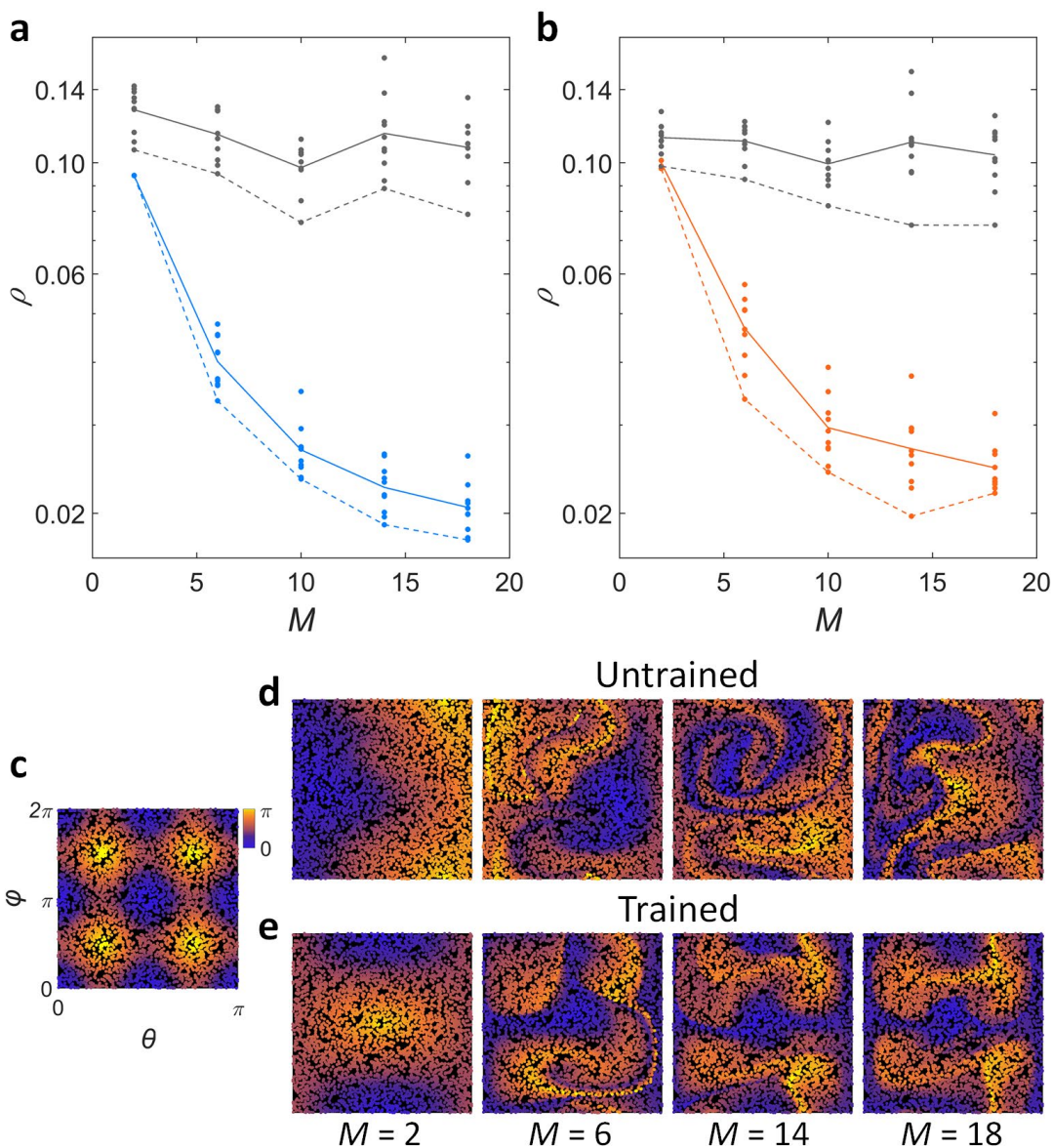

**Fig. 5. Regression with NU DNNs. a,b,** $M$-dependency of the cost functions estimated with **a,** training and **c,** test data. Colored (blue in **a** and orange in **b**) and gray plots represent the results of the trained and untrained networks, respectively. Points, solid lines, and dashed lines indicate the cost function values for each realization in the gene pool, their average, and the minimum case, respectively (Supplementary Note S7). **c-e,** Regression results for different $M$ values: the original function (**c**), the untrained network (**d**), and the trained network (**e**). For clarity, 5,000 test data points are applied in (**c-e**). The steady-state solutions in Fig. 1 are used.

**Discussion**

In conclusion, we proposed NU deep learning based on programmable photonic platforms. By utilizing norm-preserving programmable NU(1) operations, we develop an NU(2) gate as a building block for NU circuits. In a spinor circuit example, we demonstrated that our design meets the requirements for DNN hardware in terms of expressivity—complexity and coverage of mapping. We evaluated the performance of the circuit on a regression problem, employing a mutative neuroevolutionary learning [43,44]. Considering the modelling of quantum circuits as isolated systems [23] and the advantages of unitary DNNs in mitigating gradient explosion and vanishing issues [24], the proposed platform presents a promising hardware for energy-efficient and high expressive deep learning accelerators. For practical implementation and advanced functionalities, further studies are expected on intrinsic loss, more efficient training methods, customization of nonlinear functions, and the use of hysteresis for memory functions achievable with $|\Delta\omega_L| > \sqrt{3}/(2\tau)$.

**Acknowledgments:** We acknowledge financial support from the National Research Foundation of Korea (NRF) through the Basic Research Laboratory (No. RS-2024-00397664), Innovation Research Center (No. RS-2024-00413957), Young Researcher Program (No. 2021R1C1C1005031), and Midcareer Researcher Program (No. RS-2023-00274348), all funded by the Korean government (MSIT). This work was supported by Creative-Pioneering Researchers Program and the BK21 FOUR program of the Education and Research Program for Future ICT Pioneers in 2024, through Seoul National University. This work was also supported by the 2024 Advanced Facility Fund of the University of Seoul for Xianji Piao. We also acknowledge an administrative support from SOFT foundry institute. A provisional patent application (KR Prov. App. 10-2024-0179951) has been filed by Seoul National University. The inventors include S.Y., X.P. and N.P. The application, which is pending, contains proposals of nonlinear unitary circuits for deep learning.

**Fig. 1. NU(1) photonic units**. **a,** Schematic of the unit. The boxes describe the setups for $\Delta\omega_L$ and $\Delta\omega_{NL}$. Although $\Delta\omega_{NL}$ is depicted as spatially confined for clarity, the effect occurs throughout the resonator when a homogeneous material is used. **b,c,** Phase shifts $\xi(|\psi_+|^2, \Delta\omega_L)$ in linear (**b**) and nonlinear (**c**) operations. The colored surfaces in (**b,c**) and black dots in (**c**) represent the analytical (Supplementary Note S2) and numerical (Supplementary Note S3) results, which are excellently matched. Blue dashed lines represent $|\psi_+|^2 = 0.1005$ and $0.8995$. Red dashed lines represent $\Delta\omega_L = \pm 0.5161\omega_0/\tau$. $\omega_0 = 1$ and $\tau = 2{,}000/\omega_0$.

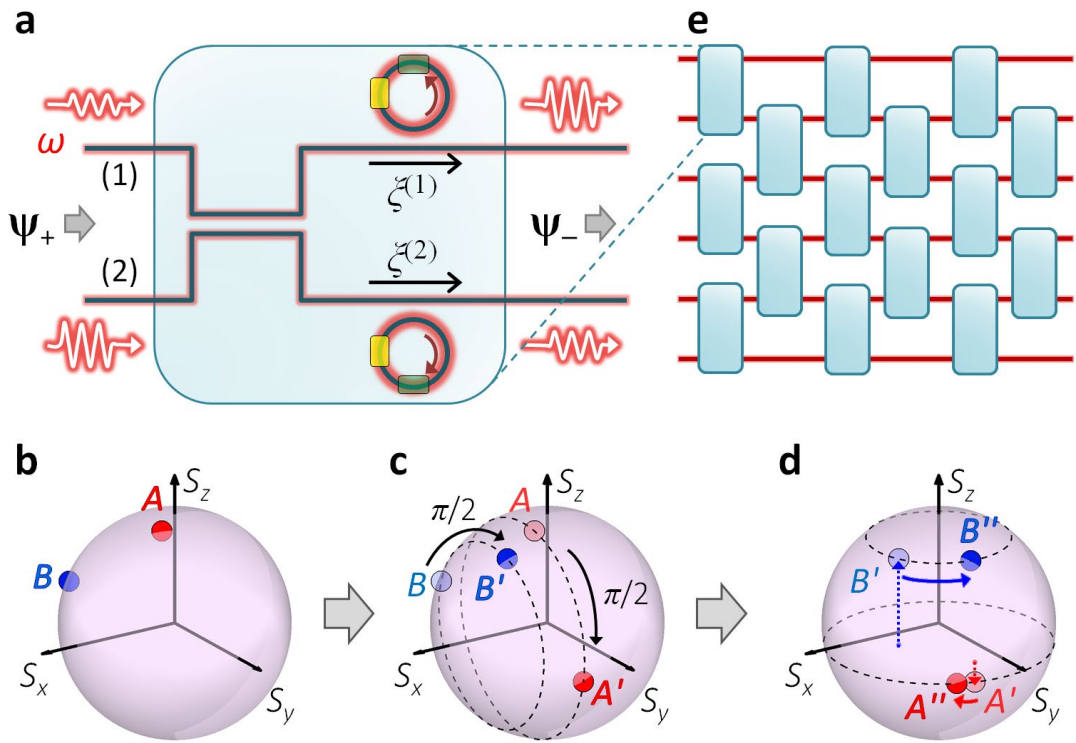

**Fig. 2. NU(2) photonic gates and circuits**. **a,** Schematic of the gate. The gate imposes an NU operation on $\psi_+$, resulting in $\psi_-$. **b-d,** Examples of the NU(2) operations: the input states $A$ and $B$ (**b**), the intermediate states $A'$ and $B'$ after the MZI (**c**), and the final states $A''$ and $B''$ (**d**). $S_{x,y,z}$ denote the spinor observables. The dashed arrows in **d** represent the $S_z$ component. All the parameters are the same as those in Fig. 1c. **e,** Schematic of an NU(6) circuit.

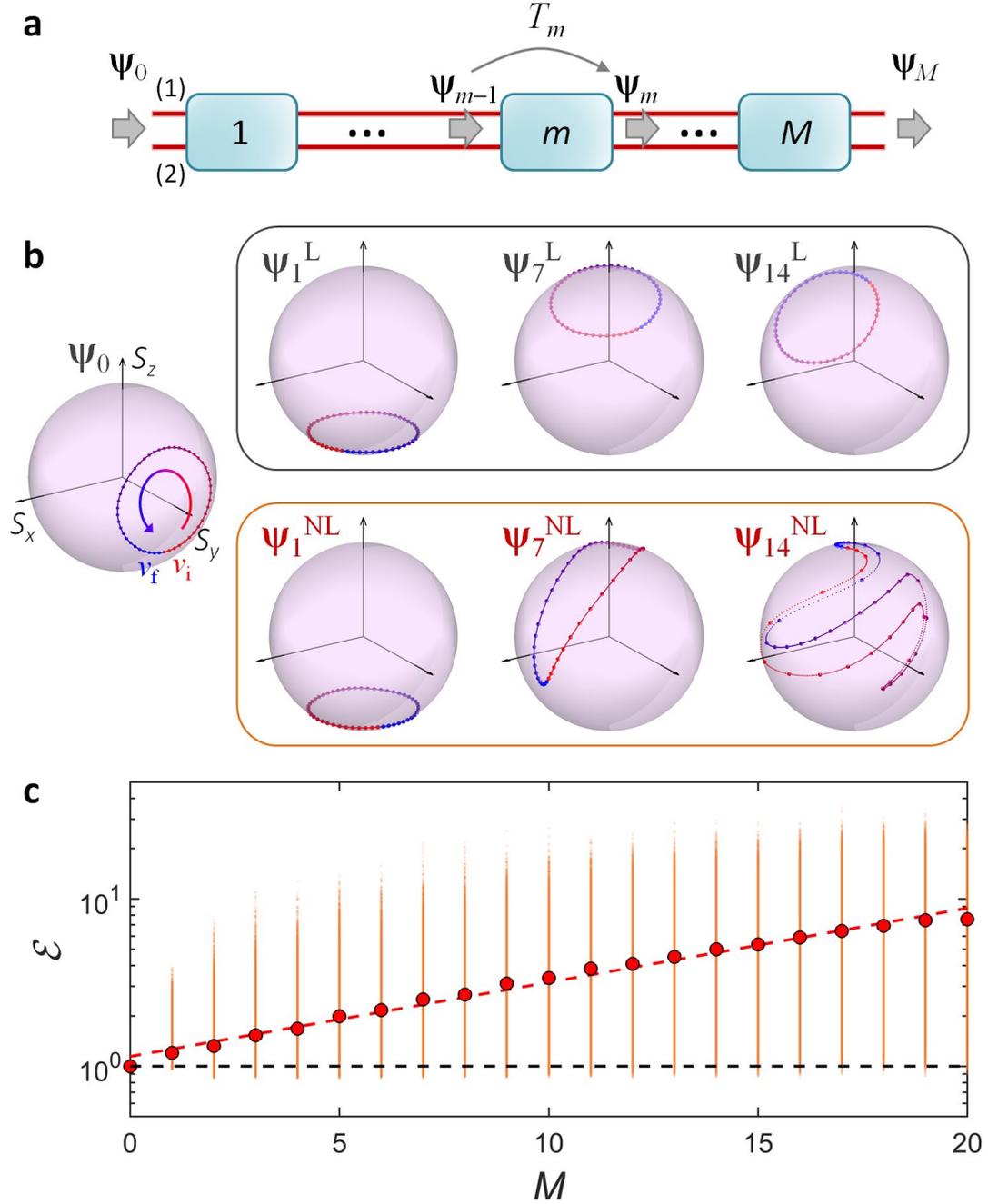

**Fig. 3. Mapping complexity**. **a,** Schematic of the NU(2) circuit used to examine expressivity. $T_m$ denotes the unitary operation achieved by the $m$-th gate. **b,** Examples of trajectory evolution through linear (top panel) and nonlinear (bottom panel) circuits for the same input trajectory (left panel) at $M = 1$, $7$, and $14$. The parameters $v_i$ and $v_f$ for Eq. (3) are indicated in red and blue, respectively. **c,** Trajectory growth as a function of $M$. Orange points represent $2 \times 10^4$ realizations: $10^3$ random system configurations according to Eq. (2), and 20 different input trajectories for each system. Red circles depict the ensemble average of $\mathcal{E}$. A black dashed line represents linear circuits: $\mathcal{E} = 1$. The steady-state solutions in Fig. 1 are used for **b** and **c**.

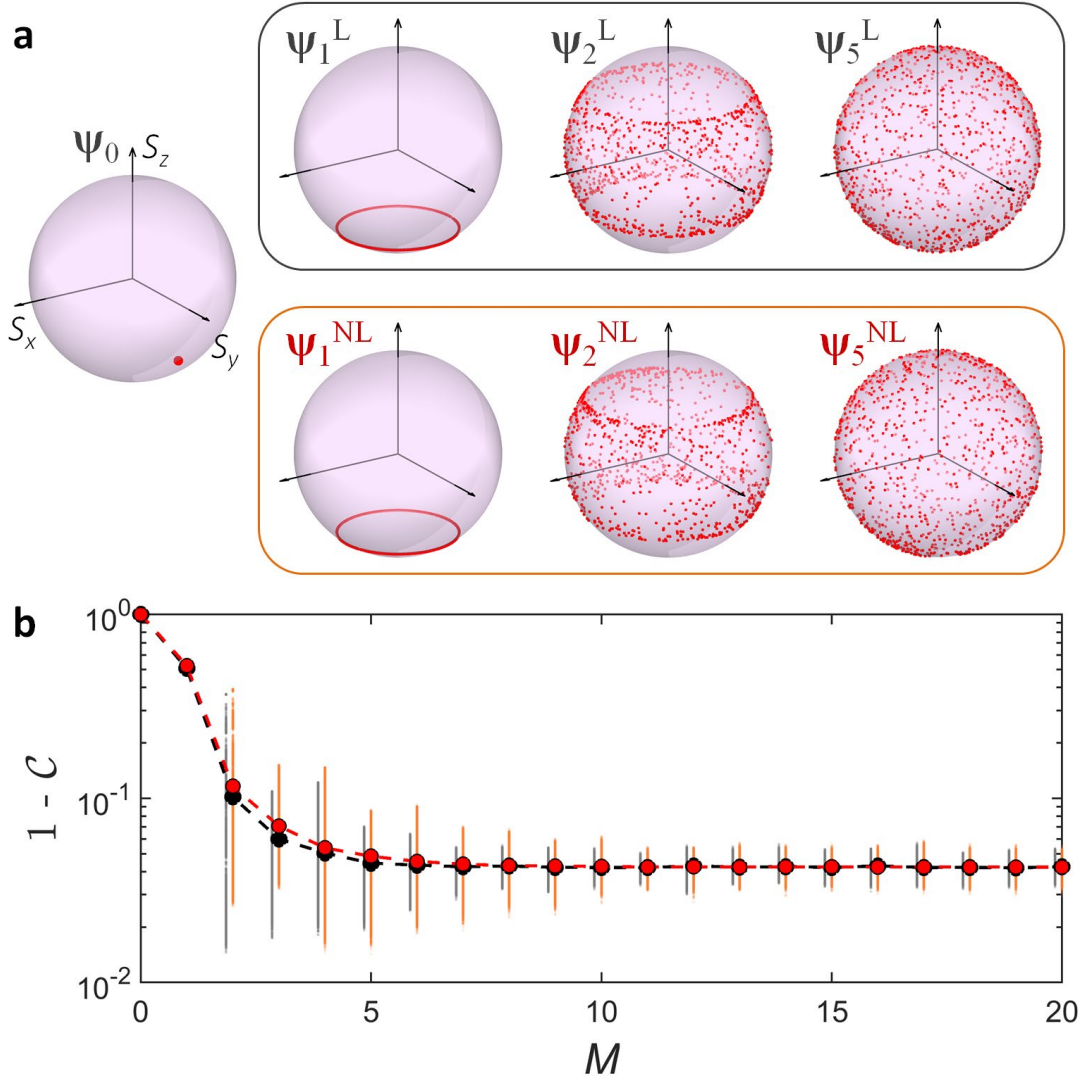

**Fig. 4. Mapping coverage**. **a,** Examples of resulting output states through linear (top panel) and nonlinear (bottom panel) circuits for an input state (left panel) at $M = 1, 2,$ and $5$. **b,** Coverage gap $1 - \mathcal{C}$ as a function of $M$ in linear (gray and black) and nonlinear (orange and red) circuits. Both gray and orange points represent $2\times10^4$ realizations from different inputs $\psi_0$: $100 \times 200$ discretization of the $\theta$–$\varphi$ input space. Both black and red circles depict the ensemble averages of $1 - \mathcal{C}$. The results of the linear circuits are slightly offset from integer $M$ values for clarity. The steady-state solutions in Fig. 1 are used.

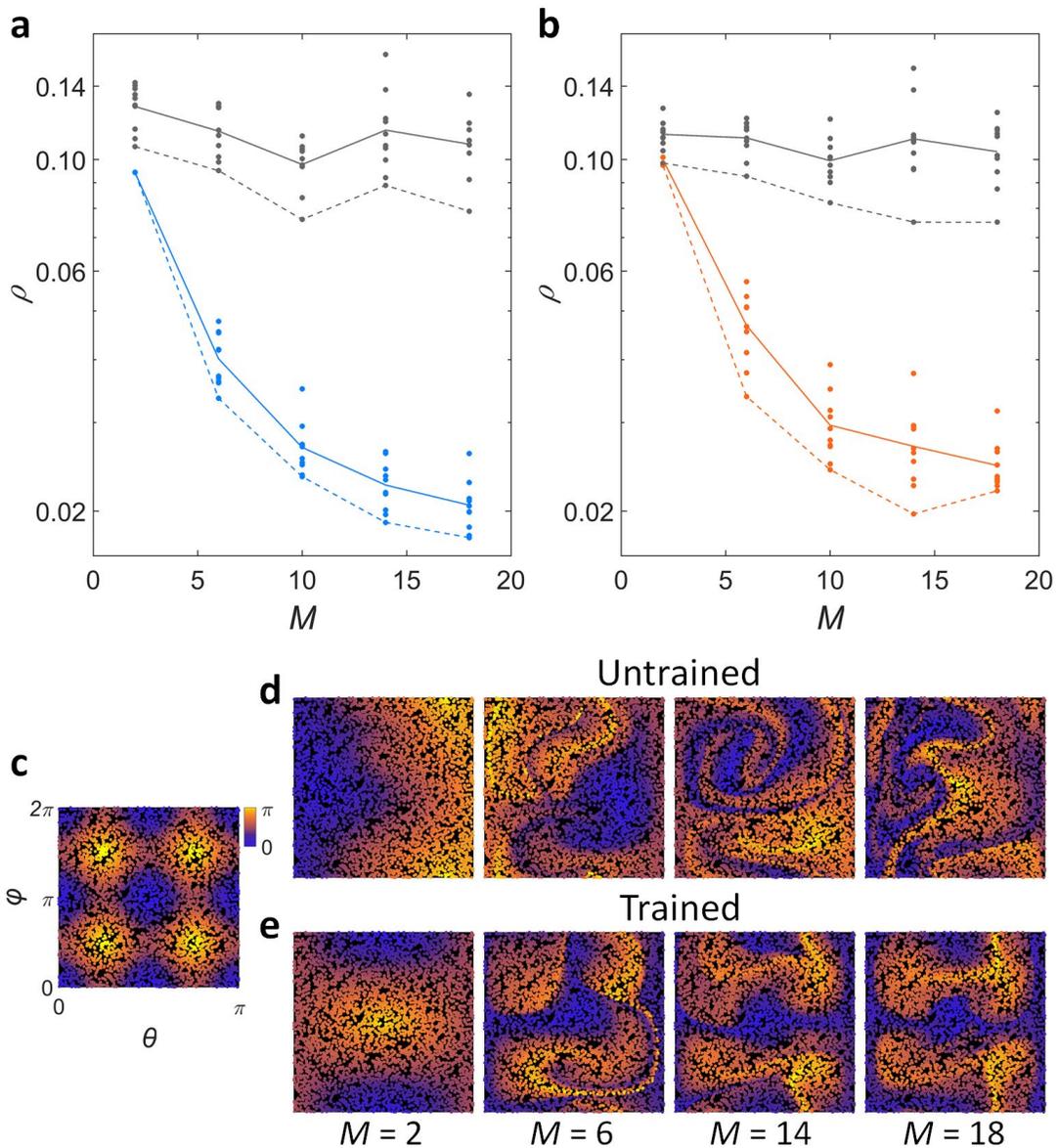

**Fig. 5. Regression with NU DNNs. a,b,** $M$-dependency of the cost functions estimated with **a,** training and **c,** test data. Colored (blue in **a** and orange in **b**) and gray plots represent the results of the trained and untrained networks, respectively. Points, solid lines, and dashed lines indicate the cost function values for each realization in the gene pool, their average, and the minimum case, respectively (Supplementary Note S7). **c-e,** Regression results for different $M$ values: the original function (**c**), the untrained network (**d**), and the trained network (**e**). For clarity, 5,000 test data points are applied in (**c-e**). The steady-state solutions in Fig. 1 are used.

# Supplementary Information for "Nonlinear Unitary Circuits for Photonic Neural Networks"

Sunkyu Yu[1†], Xianji Piao[2§], and Namkyoo Park[3*]

[1]Intelligent Wave Systems Laboratory, Department of Electrical and Computer Engineering, Seoul National University, Seoul 08826, Korea

[2]Wave Engineering Laboratory, School of Electrical and Computer Engineering, University of Seoul, Seoul 02504, Korea

[3]Photonic Systems Laboratory, Department of Electrical and Computer Engineering, Seoul National University, Seoul 08826, Korea

E-mail address for correspondence: [†]sunkyu.yu@snu.ac.kr, [§]piao@uos.ac.kr, [*]nkpark@snu.ac.kr

**Note S1. Linear operations of NU(1) units**

**Note S2. Nonlinear operations of NU(1) units at steady state**

**Note S3. Temporal dynamics of NU(1) units**

**Note S4. Nonlinear expressivity**

**Note S5. Coverage of the output space**

**Note S6. Encoding and measurement in the experimental setup**

**Note S7. Neuroevolutionary learning with simulated annealing**

**Note S1. Linear operations of NU(1) units**

For the NU(1) photonic unit depicted in Fig. 1a of the main text, the nonlinear temporal coupled mode theory is formulated as follows [1]:

$$\frac{d\mu}{dt} = i\left(\omega_0 + \Delta\omega_L + \Delta\omega_{NL} + \frac{i}{\tau_{in}} + \frac{i}{2\tau}\right)\mu + \sqrt{\frac{1}{\tau}}\psi_+,$$

$$\psi_- = \psi_+ - \sqrt{\frac{1}{\tau}}\mu, \tag{S1}$$

where $\psi_+$ and $\psi_-$ denote the incident and outgoing waves, respectively, $\mu$ represents the counterclockwise travelling-wave resonance mode, $\omega_0$ denotes the reference resonance frequency, $\Delta\omega_L$ and $\Delta\omega_{NL}$ represent resonance frequency shifts obtained from linear and nonlinear processes, respectively, and $\tau_{in}$ and $\tau$ are the lifetimes of the resonance mode due to intrinsic and coupling losses, respectively. As an ideal condition, we assume negligible intrinsic loss, given by $\tau_{in} \gg \tau$.

Consider the linear operation of the NU(1) unit with $\Delta\omega_{NL} = 0$. At the harmonic incident frequency $\omega = \omega_0$, Eq. (S1) with $d\mu/dt = i\omega_0\mu$ yields the following relationship:

$$\psi_- = -\frac{1 + 2i\Delta\omega_L\tau}{1 - 2i\Delta\omega_L\tau}\psi_+ = \psi_+ e^{-i\xi}, \tag{S2}$$

where $\xi$ is the phase shift given by $\xi = \pi - 2\arctan(2\Delta\omega_L\tau) + 2r\pi$ for integer $r$. Equation (S2) demonstrates that the side-coupled resonator can operate as a programmable norm-preserved phase shifter by modulating $\Delta\omega_L$ under the assumption of negligible loss. It is worth mentioning that the relationship between $\Delta\omega_L$ and $\xi$ corresponds to the nonlinear response of the optical quantity $\xi$ in relation to material variation $\Delta\omega_L$ [2-4], which supports understanding of the entire nonlinear operations in NU(1) units.

**Note S2. Nonlinear operations of NU(1) units at steady state**

When applying SPM with $\Delta\omega_{NL} = -\alpha|\mu|^2$ ($\alpha > 0$) to Eq. (S1), the steady-state condition at the harmonic incident frequency $\omega = \omega_0$ yields the following equilibrium relationship between the incident power $I_+ = |\psi_+|^2 \geq 0$ and the resonator energy $\chi = |\mu|^2 \geq 0$:

$$\chi\left[1 + 4\tau^2(\alpha\chi - \Delta\omega_L)^2\right] = 4\tau I_+, \tag{S3}$$

while the U(1) operation $\psi_- = \exp(-i\xi)\psi_+$ is maintained with the modified phase shift:

$$\xi = \pi - 2\arctan\left[2(\Delta\omega_L - \alpha\chi)\tau\right] + 2r\pi \quad \text{(integer } r\text{)}. \tag{S4}$$

Equation (S4) shows that the nonlinear operation of an NU(1) unit is governed by the resonator energy $\chi$, which is determined by Eq. (S3). By multiplying $\alpha$ to Eq. (S3), the cubic equation for $\alpha\chi$ is obtained as follows:

$$4\tau^2(\alpha\chi)^3 - 8\tau^2\Delta\omega_L(\alpha\chi)^2 + (4\tau^2\Delta\omega_L^2 + 1)(\alpha\chi) - 4\tau\alpha I_+ = 0, \tag{S5}$$

of which the solutions can be analytically derived using Cardano's formula for the depressed cubic form. Because multiple equilibria for real-valued $\alpha\chi$ can lead to hysteresis effects that hinder the deterministic operation of the device, we focus on achieving a unique real-valued solution. This requirement is satisfied with the condition of $D = (4D_0^3 - D_1^2) / (432\tau^4) < 0$, where $D$ is the discriminant of the cubic equation with the following $D_0$ and $D_1$:

$$D_0 = 4\tau^2\left(4\tau^2\Delta\omega_L^2 - 3\right), \quad D_1 = 32\tau^4\left(4\tau^2\Delta\omega_L^3 + 9\Delta\omega_L - 54\tau\alpha I_+\right). \tag{S6}$$

As shown in $D_1$, the condition of $D < 0$ depends on the incident light intensity $I_+$. However, we can also derive the input-intensity independent $D < 0$ for the unique equilibrium, which is achieved with $D_0 < 0$ that leads to the following relation:

$$-\frac{\sqrt{3}}{2\tau} < \Delta\omega_L < \frac{\sqrt{3}}{2\tau}. \tag{S7}$$

This result shows that the bounded linear modulation $\Delta\omega_L$ within the range of Eq. (S7) enables the hysteresis-free deterministic NU(1) operation.

In Eqs. (S4) and (S5), the origin of the nonlinear response between the incident light intensity $|\psi_+|^2$ and the phase shift $\xi$ can be attributed to two mechanisms. First, as discussed in Note S1, perturbations in the resonance lead to the arctangent relation between $\xi$ and $\alpha\chi$ in Eq. (S4), which corresponds to the nonlinear relationship between light and material signals within linear optics [2-4]. Second, the relationship between $\alpha\chi$ and $|\psi_+|^2$ is governed by the nonlinear cubic equation in Eq. (S5) for SPM, which lies in the realm of nonlinear optics.

**Note S3. Temporal dynamics of NU(1) units**

Although Eqs. (S4) and (S5) determine the steady-state operation of NU(1) units under the condition of Eq. (S7), their temporal dynamics require a direct time-domain analysis of Eq. (S1) with $\Delta\omega_{NL} = -\alpha|\mu|^2$. For this analysis, we employ the sixth-order Runge-Kutta (RK6) method [5] to solve Eq. (S1). RK6 calculations for NU(1) units are performed for harmonic incident waves with $\omega = \omega_0$. We compare the incident light intensities $|\psi_+|^2$ from 0 to 1 to characterize the effect of SPM nonlinearity on the convergence to the steady state. The simulation time is $100\tau$, and the temporal grid for the RK6 simulation is $T_0/100$, where $T_0 = 2\pi/\omega_0$. The linear resonance shift $\Delta\omega_L$ is set within the range specified in Eq. (S7).

Figures S1a-f show the temporal dynamics of the output intensity $|\psi_-|^2$ (Fig. 1a-c) and the phase shift $\xi$ (Fig. 1d-f) for different values of $|\psi_+|^2$ and $\Delta\omega_L$. Considering the functionality of the NU(1) unit as a nonlinear phase shifter, we evaluate the convergence time $t_{conv}$ of $\xi$. The convergence criterion is defined as $\xi$ converging to within an error of 0.1% of its average value over the last 3% of the total simulation time (colored circles for $t_{conv}$ in Figs. S1d-f). The case of $\Delta\omega_L > 0$ exhibits a stronger and highly nonlinear intensity dependence with longer $t_{conv}$ (Fig. 1f) due to the red shift nature of the SPM from the optical Kerr effect.

Figure S1g shows the two-dimensional (2D) map of $t_{conv}$ over the ranges of interest for $|\psi_+|^2$ and $\Delta\omega_L$. Notably, all time-varying solutions sufficiently converge to their steady states within $t_{conv} < 100\tau = 200Q/\omega_0$, where $Q$ denotes the coupling quality ($Q$-) factor of the resonator to the waveguide. According to the time-bandwidth product, the bandwidth of the NU(1) unit is approximately determined by $\omega_0 / 200Q$. We emphasize that such a bandwidth limitation can be substantially mitigated by restricting the allowed values of $\Delta\omega_L$, because slower convergence with larger $t_{conv}$ occurs only near the spectral boundary $\Delta\omega_L = \sqrt{3}/(2\tau)$.

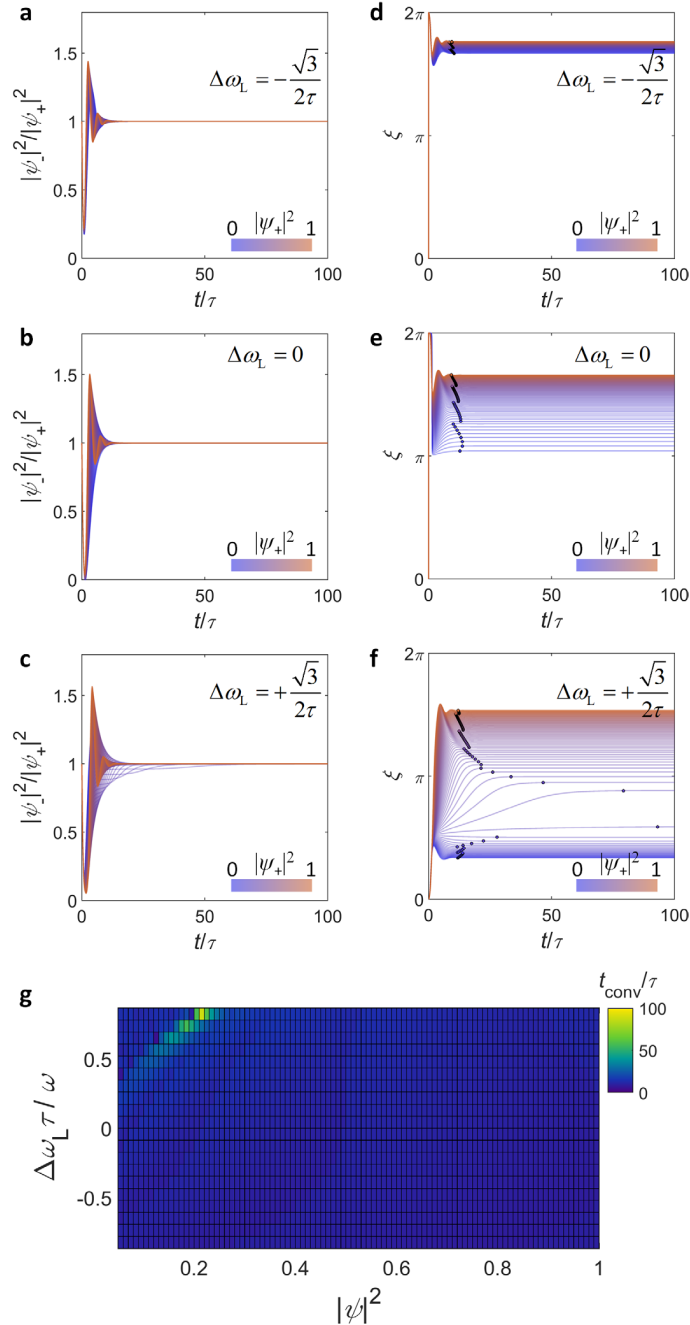

**Fig. S1. Temporal dynamics of the NU(1) units. a-c,** Output intensities $|\psi_-|^2$ and **d-f,** phase shifts for $\Delta\omega_L = -\sqrt{3}/(2\tau)$ (**a,d**), $\Delta\omega_L = 0$ (**b,e**), and $\Delta\omega_L = \sqrt{3}/(2\tau)$ (**c,f**). The dots in **d-f** indicate the time of convergence $t_{conv}$ to the steady state. **e,** The 2D map of $t_{conv}$ for the ranges of interest of $|\psi_+|^2$ and $\Delta\omega_L$ calculated by the RK6 method.

**Note S4. Nonlinear expressivity**

In our study, calculating trajectory lengths to measure nonlinear expressivity involves three steps: (i) performing circuit operations on all states of the discretized input space, (ii) preparing a set of input trajectories and characterizing the corresponding output trajectories from step (i), and (iii) calculating the lengths of the input and output trajectories using the haversine formula.

First, we discretize the Hilbert space for the input spinor states, by sampling the spherical coordinates $(\theta,\varphi)$ of the input Bloch sphere. In this discretization, $0 \leq \varphi < 2\pi$ is uniformly sampled with 200 grid points. On the other hand, the coordinate $0 \leq \theta \leq \pi$ is sampled with 100 grid points based on the cumulative distribution function of a uniformly distributed random variable on the sphere, ensuring even sampling of angles. For all input states, we perform the circuit operations using the steady-state solutions in Notes S1 and S2, for $10^3$ random system configurations according to Eq. (2) in the main text.

Second, we define a set of input trajectories $\{(\theta_l^I,\varphi_l^I)\}$, each consisting of $L$ nearby states ($l = 1, 2, \ldots, L$). The trajectory is obtained by assigning $L - 1$ movements to a random initial state $(\theta_1^I,\varphi_1^I)$. Each movement is defined as follows:

$$\left(g_\theta, g_\varphi\right) = (X, 2X), \quad X = \begin{cases} -1 & \text{if } Y < \frac{1}{3} \\ 0 & \text{if } \frac{1}{3} \leq Y < \frac{2}{3}, \quad Y \sim \text{Uniform}(0,1), \\ +1 & \text{if } Y \geq \frac{2}{3} \end{cases} \quad \text{(S8)}$$

where $g_\theta$ and $g_\varphi$ denote the grid index changes for $\theta$ and $\varphi$, respectively, considering the periodic condition of the sphere. Based on the complete mapping between the sets of input and output states from step (i), the corresponding output trajectories $\{(\theta_l^O,\varphi_l^O)\}$ are automatically determined.

To estimate the trajectory length of $\{(\theta_l,\varphi_l)\}$, denoted as $\Gamma(\{(\theta_l,\varphi_l)\})$, we apply the haversine formula to the $L - 1$ discretized cells:

$$\text{hav}(\delta_l) = \text{hav}(\theta_{l+1} - \theta_l) + \sin(\theta_l)\sin(\theta_{l+1})\text{hav}(\varphi_{l+1} - \varphi_l), \tag{S9}$$

where $\text{hav}(\zeta) = \sin^2(\zeta/2)$ is the haversine function, $\{(\theta_l,\varphi_l)\}$ and $\{(\theta_{l+1},\varphi_{l+1})\}$ denote the initial and final points of the $l$-th discretized cell ($l = 1, 2, \ldots, L - 1$), and $\delta_l$ is the central angle of the $l$-th discretized cell. For a Bloch sphere of unit radius, the length of the entire discretized trajectory is obtained as follows:

$$\Gamma(\{(\theta_l,\varphi_l)\}) = 2\sum_{l=1}^{L-1}\arctan\left(\sqrt{\frac{\text{hav}(\delta_l)}{1-\text{hav}(\delta_l)}}\right) \approx \int_{v_i}^{v_f}\left\|\frac{d\mathbf{S}}{dv}\right\|dv, \tag{S10}$$

where $\mathbf{S}(v)$ denotes the vectorial observables of the spinor states defined by $\{(\theta_l,\varphi_l)\}$, and the parameter $v$, ranging from $v_i$ to $v_f$, represents the continuous variable describing the $L$-point trajectory. By calculating $\Gamma(\{(\theta_l^I,\varphi_l^I)\})$ and $\Gamma(\{(\theta_l^O,\varphi_l^O)\})$, $\mathcal{E}$ in Eq. (3) of the main text is obtained for each pair of an input trajectory and a random system configuration.

**Note S5. Coverage of the output space**

To estimate the normalized entropy in Eq. (4) of the main text, the $\theta$-$\varphi$ output space is discretized into 12×12 cells to calculate $p_{ij}$ ($i, j$ = 1, 2, …, 12), resulting in $N_{\text{cell}}$ = 144. As in Note S4, we examine $10^3$ random system configurations for an arbitrary input $\psi_0$, achieving $10^3$ output states through circuit operations at the steady state. By classifying the obtained output states into the discretized cells, the probability $p_{ij}$ and the following coverage $\mathcal{C}$ can be calculated for the input. When the output space is perfectly covered, $p_{ij}$ = 1 / $N_{\text{cell}}$, leading to $\mathcal{C}$ = 1.

**Note S6. Encoding and measurement in the experimental setup**

As an example of deep learning using NU(2) circuits, we investigate a regression problem involving a nonconvex function: the 2D Rastrigin function [6], which is given by:

$$z = 2A + x_1^2 + x_2^2 - A\cos(2\pi x_1) - A\cos(2\pi x_2), \tag{S11}$$

where $A = 10$ and $-1 \leq x_{1,2} \leq +1$ in our example. To encode the data into spinor observables—specifically, the $(\theta_0, \varphi_0)$ coordinates of the input spinor $\mathbf{S}_0$ and the $\theta_M$ coordinate of the output spinor $\mathbf{S}_M$—we apply the following normalizations:

$$x_1 = \frac{2}{\pi}\left(\theta_0 - \frac{\pi}{2}\right), \quad x_2 = \frac{1}{\pi}(\varphi_0 - \pi), \quad z = z_{\min} + (z_{\max} - z_{\min})\frac{\theta_M}{\pi}, \tag{S12}$$

where $z_{\min}$ and $z_{\max}$ denote the predefined minimum and maximum values of $z$ for rescaling $\theta_M$.

Figure S2 illustrates the proposed experimental setup. The variables $x_1$ and $x_2$ are encoded into $\psi_0$ and its observable $(\theta_0, \varphi_0)$ via a universal SU(2) gate [7]. Through the NU(2) circuit, we obtain $\psi_M = [\psi_M^{(1)}, \psi_M^{(2)}]^T$. By measuring the light intensity of the output state through photodetection, we obtain the target output observable, as follows:

$$\theta_M = 2\arctan\left(\sqrt{\frac{|\psi_M^{(2)}|^2}{|\psi_M^{(1)}|^2}}\right). \quad (0 \leq \theta_M \leq \pi) \tag{S13}$$

The output of the 2D Rastrigin function is reconstructed by Eq. (S12) using $\theta_M$.

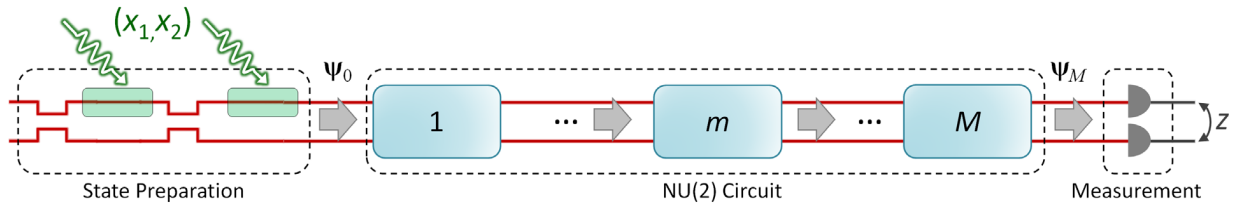

**Fig. S2. Proposed experimental setup.** The system is composed of the state preparation, NU(2) circuit, and photodetection. In the preparation stage, $x_{1,2}$ are encoded into $\psi_0$. After performing the circuit operation, we obtain $\psi_M$, which is converted into $z$ through photodetection.

**Note S7. Neuroevolutionary learning with simulated annealing**

To train the NU(2) circuit for the regression problem of Note S6, we employ neuroevolutionary learning [8,9] combined with simulated annealing [10]. In this learning process, we focus on mutative neuroevolution, which corresponds to a stochastic generalization of gradient-descent methods [9]. For training and evaluation, we prepare 1,000 data points and 200 data points, respectively, for the set $(x_1, x_2, z)$ using Eq. (S11).

The core mechanism of neuroevolutionary learning lies in the mutation of NU(2) circuits. Such mutative evolution over the evolution epochs $s = 1, 2, \ldots, s_{\max}$ is characterized by a series of perturbations in the system parameters of the NU(2) circuit, $\{\Delta\omega_{L,m}^{(1,2)}(s), \alpha_m^{(1,2)}(s)\}$. For the linear resonance shifts $\Delta\omega_{L,m}^{(1,2)}(s)$, which can have continuous values by using tunable phase shifters, we apply perturbations based on a uniform random distribution, as follows:

$$\Delta\omega_{L,m}^{(1,2)}(s+1) = \Delta\omega_{L,m}^{(1,2)}(s) + \delta\omega_i \left(\frac{\delta\omega_f}{\delta\omega_i}\right)^{\frac{s-1}{s_{\max}-1}} \text{Uniform}(-1,1), \quad \text{for all } m \quad (S14)$$

where $\delta\omega_i$ and $\delta\omega_f$ denote the initial and final magnitudes of perturbations, respectively. We set $\delta\omega_i = 0.4 \times \sqrt{3}/(2\tau)$ and $\delta\omega_f = 0.02 \times \sqrt{3}/(2\tau)$, analogous to decreasing the learning rate in the gradient descent method, which guarantees the convergence to a local minimum. On the other hand, the nonlinear coefficients can have binary values, 0 or $\alpha_0$, due to the use of a homogeneous nonlinear material. Therefore, we employ the following discretized perturbation, altering the nonlinearity of only a single gate, as follows:

$$\alpha_m^{(1,2)}(s+1) = \alpha_0 - \alpha_m^{(1,2)}(s), \quad \text{for } m \sim \text{Uniform}(1, M). \quad (S15)$$

For neuroevolutionary learning governed by the above mutative evolutions, we prepare a gene pool composed of 10 differently initialized NU(2) circuits using Eq. (2) in the main text. With the gene pool, we iteratively employ the mutations, estimate the cost functions, and determine whether to accept the mutations regarding the cost functions. To obtain the cost function for each circuit, we perform the circuit operation with the mutative system parameters from Eqs. (S14) and (S15), resulting in $z(s)$ from Note S6. The cost function is then defined by the mean square error $\rho(s) = \langle |z(s) - z_{\text{true}}|^2 \rangle_D$, where $z_{\text{true}}$ denotes the ground truth data from Eq. (S11) and $\langle \ldots \rangle_D$ denotes the ensemble average over the dataset.

To handle the nonconvex function, such as the 2D Rastrigin function in our example, we employ simulated annealing [10] in the acceptance process to search for the global optimum. The process is determined by the following acceptance probability $P(s)$:

$$P(s) = \min\left(1, \exp\left(-\frac{\rho(s) - \rho(s-1)}{T_{\text{SA}}(s)}\right)\right), \quad (s = 1, 2, \ldots, s_{\max}) \tag{S16}$$

where $T_{\text{SA}}(s)$ is the temperature, which controls the probability of accepting worse solutions to explore a broader solution space. In our study, $T_{\text{SA}}(s)$ is defined as follows:

$$T_{\text{SA}}(s) = T_{\text{i}} \left(\frac{T_{\text{f}}}{T_{\text{i}}}\right)^{\frac{s-1}{s_{\max}-1}}, \tag{S17}$$

where $T_{\text{i}} = T_{\text{SA}}(s=1)$ and $T_{\text{f}} = T_{\text{SA}}(s=s_{\max})$ denote the initial and final temperatures, respectively. In our training process, we set these hyperparameters as $T_{\text{i}} = 2 \times 10^{-4}$ and $T_{\text{f}} = 1 \times 10^{-8}$. The mutation defined by Eqs. (S14) and (S15) at epoch $s$ is accepted if:

$$P(s) \geq \text{Uniform}(0,1). \tag{S18}$$

After a sufficient number of evolutions, set to $s_{\max} = 4 \times 10^6$ in our example, we obtain the optimized NU(2) circuits for regression. We evaluate Fig. 5 in the main text using both the untrained and optimized circuits. The process is described with the pseudo-code in Algorithm S1.

**Algorithm S1. Pseudo-code for neuroevolutionary learning.** This method describes the pseudo-code for optimizing the NU(2) circuits with depth $M$ for the regression problem described in Fig. 5 in the main text and Note S6.

1:   Initialize 1,000 training data points and 200 test data points using Eq. (S11)

2:   Initialize a gene pool composed of 10 sets of $\{\Delta\omega_{L,m}^{(1,2)}(0), \alpha_m^{(1,2)}(0)\}$ ($m$ = 1, 2, …, $M$)

3:   Apply circuit operations with $\{\Delta\omega_{L,m}^{(1,2)}(0), \alpha_m^{(1,2)}(0)\}$ to the training and test input data

4:   Estimate $\rho(0)$ for both datasets

5:   **for** every $s$ where $1 \leq s \leq s_{max}$ **do**

6:        Apply mutations to $\{\Delta\omega_{L,m}^{(1,2)}(s–1), \alpha_m^{(1,2)}(s–1)\}$, obtaining $\{\Delta\omega'_{L,m}^{(1,2)}(s), \alpha'_m^{(1,2)}(s)\}$

7:        Apply circuit operations with $\{\Delta\omega'_{L,m}^{(1,2)}(s), \alpha'_m^{(1,2)}(s)\}$ for the training input data

8:        Estimate $\rho(s)$ with the training output data

9:        Estimate $P(s)$ with Eq. (S16)

10:       **if** $P(s)$ satisfies Eq. (S18)

20:            Accept mutations: $\{\Delta\omega'_{L,m}^{(1,2)}(s), \alpha'_m^{(1,2)}(s)\} \rightarrow \{\Delta\omega_{L,m}^{(1,2)}(s), \alpha_m^{(1,2)}(s)\}$

21:       **else**

22:            Reject mutations: $\{\Delta\omega_{L,m}^{(1,2)}(s–1), \alpha_m^{(1,2)}(s–1)\} \rightarrow \{\Delta\omega_{L,m}^{(1,2)}(s), \alpha_m^{(1,2)}(s)\}$

23:       **end if**

24:  **end for**

25:  Apply circuit operations with $\{\Delta\omega_{L,m}^{(1,2)}(s_{max}), \alpha_m^{(1,2)}(s_{max})\}$ to the test input data

26:  Estimate $\rho(s_{max})$ with the test output data